\documentclass[fleqn,usenatbib]{mnras}

\usepackage{newtxtext,newtxmath}

\usepackage[T1]{fontenc}
\usepackage{ae,aecompl}

\usepackage{graphicx}	
\usepackage{amsmath}	
\usepackage{amssymb}	

\newcommand{\hei}{He {\sc I}}
\newcommand{\hii}{H {\sc II}}
\newcommand{\ha}{H$\alpha$}
\newcommand{\hb}{H$\beta$}
\newcommand{\oiii}{[O {\sc III}]}
\newcommand{\nii}{[N {\sc II}]}
\newcommand{\sii}{[S {\sc II}]}

\title[Emission-line Profile of and Variability in 3C 445]{Profile of and Variability in Double-Peaked Balmer Emission Lines in 3C 445}

\author[S. Zhang et al.]{
Shaohua Zhang,$^{1}$\thanks{E-mail: zhangshaohua@pric.org.cn}
Hongyan Zhou,$^{1,2,3}$
Xiheng Shi,$^{1}$
Bo liu,$^{1,2,3}$
and Xiang Pan$^{1}$
\\
$^{1}$SOA Key Laboratory for Polar Science, Polar Research Institute of China, 451 Jinqiao Road, Shanghai 200136, China\\
$^{2}$CAS Key Laboratory for Research in Galaxies and Cosmology, University of Sciences and Technology of China, Hefei, Anhui 230026, China\\
$^{3}$School of Astronomy and Space Science, University of Science and Technology of China, Hefei 230026, China}

\date{Accepted XXX. Received YYY; in original form ZZZ}

\pubyear{2019}

\begin{document}
\label{firstpage}
\pagerange{\pageref{firstpage}--\pageref{lastpage}}
\maketitle

\begin{abstract}
We extract the multiple-epoch Balmer-line profiles of the heavily obscured quasar 3C 445 from the spectral curves in the literature, and analyze the emission-line profiles of the H$\alpha$ and H$\beta$ lines and the profile variability in the H$\alpha$ line in the large time interval of more than three decades. The profile comparison between the H$\alpha$ and H$\beta$ lines shows that both Balmer lines share the profile with the same form, while the blue system of the H$\beta$ line is seriously weaker than that of the H$\alpha$ line. Moreover, the blue system of the H$\alpha$ line suddenly disappeared completely and then did not appear again, however the other two components did not exhibit significant variation in the velocity or the amplitude. These findings suggest that the blue system of 3C 445, as with SDSS J153636.22+044127.0 and its analogs, is probable the result of the shock-heated outflowing gases. The observation angle of almost edge-on which the previous studies suggested can easily  produce the high-speed and high-temperature shock in the collision between the massive outflow and the inner surface of the dusty torus.
\end{abstract}

\begin{keywords}
galaxies: active -- quasars: emission lines --  quasars: individual (3C 445)
\end{keywords}



\section{Introduction}

The emission-line system of quasars provides an effective way to understand the structure and gas properties of active galactic nuclei (AGNs). It has been traditional view that the broad-line emission originates in inner high-density clouds orbiting the central ionizing radiation source, and the origin of narrow-line is the extended low-density materials at larger scales, which are called the broad-line region (BLR) and the narrow-line region (NLR), respectively, in the unified model (Antonucci 1993). Moreover, with the increase in spectral observation data and the deepening of research on emission-line profiles, researchers have been cognizant of that other mechanisms, such as the accretion disk (e.g., Collin-Souffrin et al. 1980; Chen et al. 1989; Oke 1987; Halpern 1990), the outflowing of gas at the different scales (e.g., Leighly 2004; Wang et al. 2011; Marziani et al. 2013; Liu et al. 2016; Zhang et al. 2017a), and even the dusty torus (e.g., Li et al. 2015, 2016), can seriously affect the broad-line profile and  produce some unique emission-line profiles.

To investigate the origins of the unique  emission-line profiles,  follow-up observations and other high-/low-ionization lines extending to the infrared (or/and ultraviolet) bands, assisted by kinetic and photoionization simulations, can effectively diagnose the true source of the anomalous emission-line system. For example, it was once considered that in SDSS J153636.22+044127.0 and its analogs, the blue system of the double-peaked hydrogen Balmer lines arises from the broad-line region of a secondary supermassive black hole (Boroson \& Lauer 2009; Tang \& Grindlay 2009); however,  new evidences from the follow-up optical/near-infrared spectral observations, i.e., the profile invariance in the time interval of 10 years   and the absence of the blue system in the \hei\  $\lambda$10830  emission-line profile, suggests that these peculiar emission-line profiles are is probably related to the shock-heated outflowing gases (Zhang et al. 2019).

3C 445 ($z=0.0562$; Osterbrock et al. 1976) is also a widely studied broad-line radio galaxy with double-peaked Balmer emission lines, in which irregular, non-symmetric profiles were suggested to result from mass motions of the ionized gas in a relatively small number of ``clouds'' or ``streams'' at one time (Osterbrock et al. 1976). Indeed, the Balmer lines in 3C 445 are more complex than most ordinary double-peaked emitters (see the samples in Eracleous \& Halpern 2003 and Strateva et al. 2003, 2004). In the following three decades, there have been several spectral observations of its emission-line profiles (e.g., Crenshaw et al. 1988; Eracleous \& Halpern 1993; Corbett et al. 1998;  Buttiglione et al. 2009). These spectral archives would provide an approach to exploring the origins of the emission-line components of the Balmer lines with the unique flux ratios and different variations.

\section{Spectral Data of the Balmer Lines}

\begin{figure*}
\includegraphics[width=2.0\columnwidth]{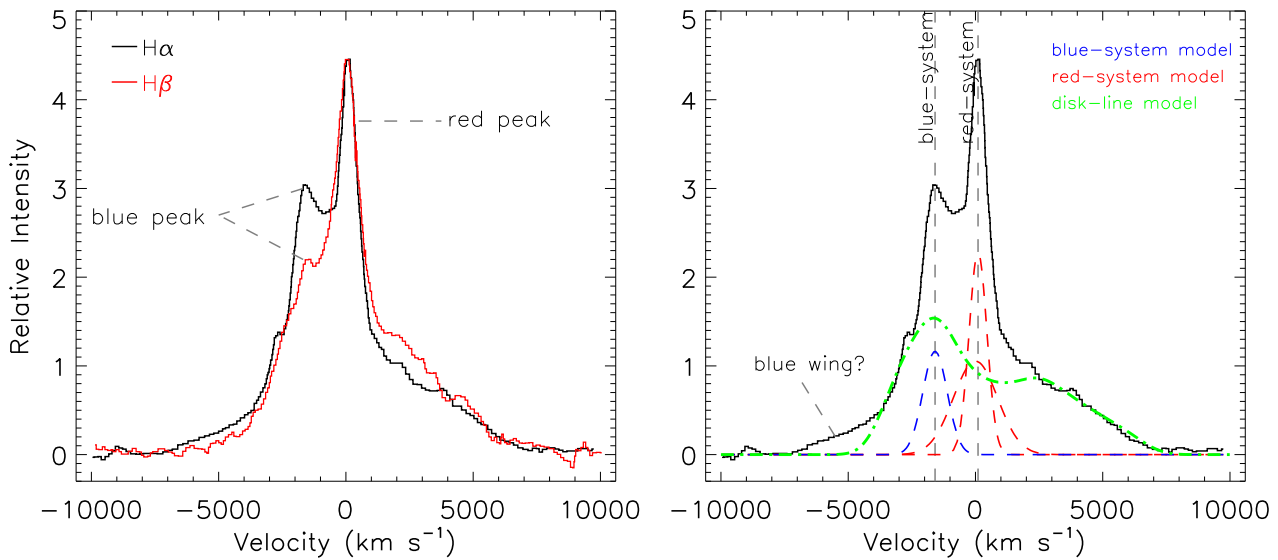}	
\caption{Left: Balmer-line profile comparison of 3C 445 (scanned in 1974). Two Balmer lines share the same profile within the observational accuracy except the blue peak. Intriguingly, the blue peak of the H$\beta$-line profile at $\sim 1600~\rm km~s^{-1}$ is significantly lower than that of H$\alpha$. The emitting gases of the blue peak  are probable with different intrinsic gas/dust properties  from the other emission-line components.
Right: H$\alpha$-line profile and its decomposition of 3C 445. The black line is the continuum-removed spectrum from  Osterbrock et al. (1976), the green dash-dotted  line is the best-fit disk-line model, the left blue dashed line is the Gaussian fit to the blue peak (marked as  the ``blue system''), the central red dashed lines are the two Gaussians fit to the red peak (marked as the ``red system'').}\label{fig:f1}
\end{figure*}

As a famous double-peaked broad-line source, there are many spectral observations of 3C 445 in the literature. In this work, we  choose only five optical spectral observations  in the large time interval of more than three decades, which all provide high-quality  emission-line profiles of the H$\alpha$ line that can be used to check the  possible evolution of the broad emission lines.

Two scans of 3C 445, 64 minutes in blue and 48 minutes in red spectral regions, respectively, were firstly taken with  the 120 inch telescope at Lick Observatory on 1974 July 22${\rm nd}$ and September 9${\rm th}$, and the summed scans were compared with the other three broad-line radio galaxies in Figure 1 of Osterbrock et al. (1976). 
In Figure 6 of Osterbrock et al. (1976), the emission-line profiles of Balmer lines, with the blended \nii, \sii, and \oiii\ lines removed, are displayed. We extract the relative intensities of the H$\alpha$ and H$\beta$ lines from the curves in the figure  using a semiautomated tool called ``WebPlotDigitizer''\footnote{https://automeris.io/WebPlotDigitizer/}.

The first follow-up spectrum of 3C 445 has already presented the large-amplitude variability in the H$\alpha$ line within the velocity range of the blue system. It was obtained with the Ohio State University Image Dissector Scanner (IDS) on the 1.8 m Perkins reflector at Lowell Observatory on 1986 November 5${\rm th}$.
An exposure of 5400 seconds was taken in the red  region and the 600 lines mm$^{-1}$ grating provide a wavelength range of 2000 \AA\ centered on approximately 6800 \AA.
The new H$\alpha$-line profile is extracted from the curve in the last panel of Figure 1 of Crenshaw et al. (1988).

Five years later, two observations (with exposure times of 2200 and 2400 seconds)
were carried out using the 2.1 m telescope and GoldCam CCD spectrograph at the Kitt Peak National Observatory on 1991 June 20${\rm th}$ and 21${\rm st}$. The spectral  range was  $6330-7440$ \AA\ to include H$\alpha$ and its broad wings with a spectral resulation of 3.7 \AA.
Two H$\alpha$-line profiles are extracted from the curves in Figure 1-(g) of Eracleous \& Halpern (1993), and there is no evidence for profile variability  on the time scale of a few days. Thus we combine two H$\alpha$-line profiles for the following analysis.

The third follow-up spectrum is taken from the spectropolarimetric observations  on 1995 June 6${\rm th}$ and 7${\rm th}$. Three polarization spectra were obtained on the red arm of the ISIS dual-beam spectrograph on the 4.2 m William Herschel Telescope at the Observatorio del Roque de los Muchachos with  exposure times of 2000, 2000 and 800 seconds. A 316 line mm$^{-1}$ grating gave a wavelength range of 1500 \AA\ and a dispersion of  approximately 1.5 \AA\ pixel$^{-1}$.
Since the last observation has a significantly different continuum polarization position angle from those of the first two observations (which are consistent), an average spectrum was created by combining the first two observations, as shown in Figure 1-(c) of Corbett et al. (1998). We extract the total flux spectrum.

The fourth follow-up spectrum is obtained with the 3.58 m optical/infrared Telescopio Nazionale Galileo (TNG) with the DOLORES spectrograph on 2007 August 6th (Buttiglione et al. 2009). The chosen long-slit width is 2 arcsec. An exposure of 500 seconds was taken with the VHR-R ($6100 - 7800$ \AA) grism with a resolution of $\sim 5$ \AA. The H$\alpha$-line spectrum is downloaded from the NASA/IPAC Extragalactic Database (NED).
For the third and the fourth follow-ups, the local continua were subtracted from the observed spectra to obtain the  H$\alpha$-line profiles. The local continuum of the H$\alpha$ regime is estimated from two continuum windows ([6200, 6250] \AA\ and [6800, 6850] \AA) in the form of a linear function.

\section{Profile Analysis }
\subsection{Profile Comparison of the Bamler Lines and Unusual Blue System}
In the left panel of Figure \ref{fig:f1}, we compare the emission-line profiles of the H$\alpha$ and H$\beta$ lines. As Osterbrock et al. reported, the observed emission-line profile of the H$\beta$ line has the same form as the H$\alpha$-line profile within the observational accuracy. However, intriguingly, the blue peak of the H$\beta$-line profile at $\rm \sim - 1600\ km\ s^{-1}$ is significantly weaker than that of the H$\alpha$-line profile.
It is more likely that the Balmer profiles in 3C 445 are more complex than most ordinary double-peaked emitters, which is explained as an relativistic accretion disk plus one set of AGN's broad/narrow emission lines at the zero velocity (e.g., Halpern 1990; Eracleous \& Halpern1994).

Carefully analyzing the emission-line profiles of the \ha\ and \hb\ lines, we found that the profiles can be decomposed to be a double-peaked baseline  plus two emission-line systems, which are shown by the green dash-dotted line and marked as the ``blue system''  and the ``red system'' in the right panel of Figure \ref{fig:f1}, respectively. In particular, the double-peaked baseline is modeled using an axisymmetric Keplerian disk-line model from Chen \& Halpern (1989), the disk inclination is $i\sim 48^\circ$, the inner/outer radii are $r_{\rm 1}\sim 1400\ r_{\rm G}$ and $r_{\rm 2}\sim 13,000\ r_{\rm G}$, the velocity dispersion is $\sigma = 600\rm\ km\ s^{-1}$, and the index of the surface emissivity power law is $q=2.5$. We should know that the first three parameters in the disk model, i.e., the inclination angle and the inner and outer radii, are degenerate to some degree, and the larger inclination angle would lead to the larger inner radius (see Figure 3 of Tang \& Grindlay 2009), therefore, the bestfit parameters are only good as a rough estimate.
In the right panel of Figure \ref{fig:f1}, the shoulder-like peak is obscured by the blue system, plus, there is a blue wing component which cannot be fit well by an axisymmetric Keplerian disk model, as shown obviously in the H$\alpha$ profile, which was also found in SDSS J153636.22+044127.0 (Tang \& Grindlay 2009).
Moreover, the blue and red systems are portrayed by one single Gaussian (blue dashed line), and two Gaussians (red dashed lines), and the velocity shifts of the blue system and the broad and narrow components of the red system are $-1588$, 0, and $95\rm\ km\ s^{-1}$, and their full width at half maximum (FWHM) values are 1132, 2051, and $825\rm\ km\ s^{-1}$, respectively.

Indeed, the Balmer profile properties (the whole emission-line profile includes the blue/red systems with large velocity offset and a double-peaked baseline,  and the flux ratio of the blue system is significant different from those of other two emission-line components) of 3C 445 are very similar to those of SDSS J153636.22+044127.0 and its analogs reported by Zhang et al. (2019). The two significant differences are that the emission strength of the blue system of 3C 445 is weaker, 
and the velocity offset between the blue and red systems of 3C 445 is also smaller than those of these analogs. Generally, the double-peaked baseline is the disk-line component of the accretion disk around the black hole,  the red system is the emission component originating from the normal emission-line regions.  In Zhang et al. (2019), the blue system is considered to be probably related to the shock-heated outflowing gases. Similarly for 3C 445, large disagreement between the \ha\ and \hb\ blue systems and rough consistency at other velocities imply that the emitting gases of the blue system are probable with different intrinsic gas/dust properties (e.g., the  abnormal intrinsic \ha/\hb\ ratio, and extra extinction)  in the nuclear region from the other two emission-line components.

\begin{figure}
	\includegraphics[width=\columnwidth]{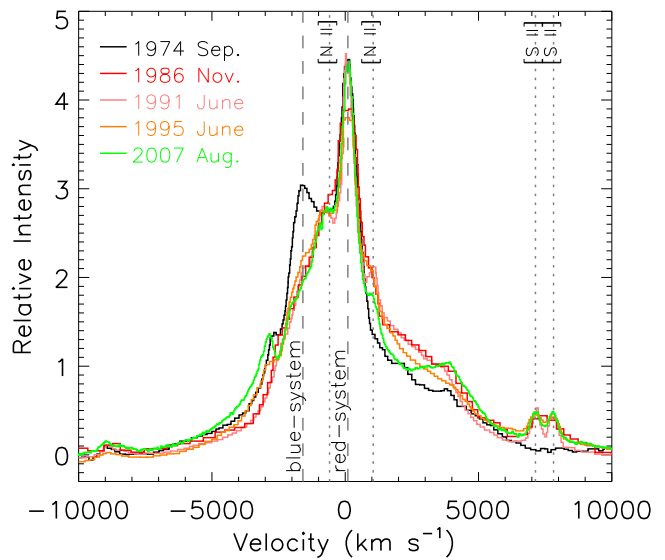}
    \caption{H$\alpha$-line profile comparison of 3C 445 from 1974 to 2007.  The H$\alpha$ lines in the literature share the same profile within the observational accuracy except the blue system. The blue system of H$\alpha$ suddenly disappeared between 1974 and 1986, and did not appear again in the next two decades. The different  variations in the three emission-line components are probable to be related to the origins of the complex Balmer-line profile. }
    \label{fig:f2}
\end{figure}

\subsection{Blue System Variability in the H$\alpha$ Line}
In Figure \ref{fig:f2}, we present the H$\alpha$-line profiles of 3C 445 in the literature in the velocity space. Each follow-up profile of  the H$\alpha$ line is multiplied by a factor to match the red peak  of the earliest profile. The narrow lines blended in the H$\alpha$-line profile do not interfere in the comparison of broad-line profiles; thus the \nii\  and \sii\ lines are not removed (in the profile of 1974 observation these lines have been removed by Osterbrock et al. 1976), and we mark these narrow lines with dotted lines in the figure. Through the comparison of the emission-line profiles, the blue system of the H$\alpha$ line suddenly disappeared completely between 1974 and 1986, but in the next two decades after 1986, the profile of the H$\alpha$ line has no significant variation in the velocity or the amplitude. In detail, the three components contained in the Balmer-line profiles exhibit different variations, 
i.e., the absent blue system, and other two emission components without significant variation. We also checked other emission-line profiles of the H$\alpha$ line in the literature (e.g., Cohen et al. 1999; Keel et al. 2005; Jones et al. 2009). These spectra are observed in the 1990s and beyond, and they have low spectral resolution or their observed times are very close to one of the spectra we have presented in Figure \ref{fig:f2}. As we expected, the blue system is also absent in these H$\alpha$ profiles. These differences in the variability of the three emission-line components are probable to be related to the origins of the complex Balmer-line profile suggested in the last section.
Of course, there are also people who think that the disappearance of the H$\alpha$ blue system may be the profile change of the double-peaked disk line, which generally represents the variations of relatively large-scale accretion disks in AGNs (e.g., Lewis et al. 2010 and references therein). In this situation, the H$\alpha$ blue system in the 1974 observation is part (i.e., the blue shoulder) of the disk-line component. However, that cannot explain the profile difference of the Balmer lines, the maximum relative intensity of the H$\beta$ blue peak is only $\sim70\%$ of the H$\alpha$ blue peak. Moreover, the disk-line model needs the extremely small inclination angle to match the high blue peak of the H$\alpha$ line, and the broad extended wings of the H$\alpha$-line profile cannot be explained at all.

\section{SUMMARY AND DISCUSSION}
In this work, we extract the  multiple-epoch  Balmer-line  profiles of 3C 445 from the curves in  the literature, and  analyze the emission-line profiles of the H$\alpha$ and H$\beta$ lines and the profile variability in the H$\alpha$ line in the large time interval of more than three decades.
The comparison between the H$\alpha$ and H$\beta$  profiles  in the first observation shows that  both  lines share the  emission-line profile with the same form, while the blue system of the H$\beta$ line with a blueshifted velocity of $\sim 1600\rm \ km\ s^{-1}$ is much lower than that of the H$\alpha$ line.
Intriguingly, in the next three decades following Osterbrock et al.'s spectral observation, the double-peaked baseline and the red system (at the zero velocity) of the H$\alpha$ line in 3C 445 have no significant variation in the velocity or the amplitude, however, the H$\alpha$ blue system suddenly disappeared completely and then did not appear again. The differences in the flux ratio and the profile variability between the three emission-line components suggest that the origin of the unique Balmer-line profile of 3C 445 is complex and  that the three  components emit from the different gas clouds in the nuclear region of 3C 445.

Generally, the double-peaked baseline is thought to be the disk-line component of the accretion disk around the black hole, and the red system is the emission component originating from the normal emission-line regions. In previous studies, the extra blue system in the broad emission lines had been considered to be coming from the BLR around the secondary black hole in a binary black hole system (e.g., Boroson \& Lauer 2009; Tang \& Grindlay 2009; Decarli et al. 2010), or those are closely related to the outflowing gases (e.g., Leighly 2004; Liu et al. 2016; Zhang et al. 2017b, 2019). In the binary black hole model, the periodic changes in the velocity offset between the blue and red systems and flux amplitude of the blue system are expected, obviously, this binary hypothesis is not entirely consistent with the observed facts of 3C 445. The blue system of the  H$\alpha$-line profile disappeared suddenly in the 12 years after the first spectral observations, and then  never appeared  as scheduled in the next two decades.

\begin{figure}
	\includegraphics[width=\columnwidth]{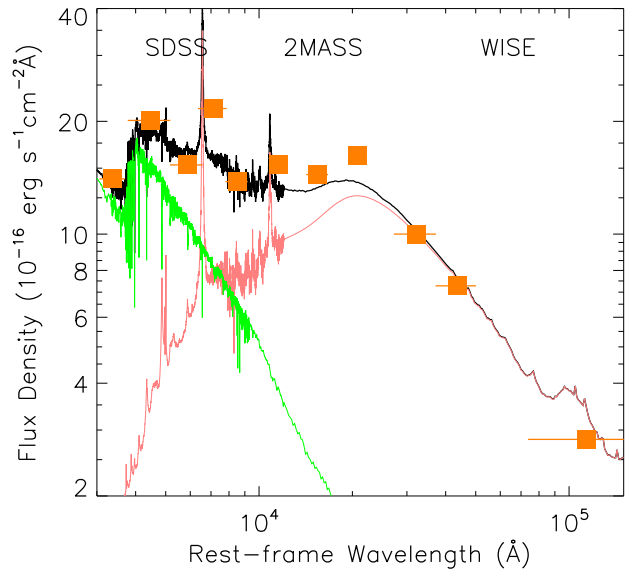}
    \caption{Broadband SED of 3C 445 from optical to middle-infrared by orange squares. The reddened quasar composite, the `$cst\_6Gyr\_z008$' galaxy template and their sum are shown by pink, green, and black curves. }
    \label{fig:f3}
\end{figure}

In Table 2 of Osterbrock et al. (1976), the combined intensities of the H$\alpha$ and H$\beta$ lines measured from the first spectrum were listed.
The rough Balmer decrement of 3C 445 ($\rm H\alpha/H\beta=47.2/4.97$) is much steeper than the recombination decrement. 
In Veilleux \& Osterbrock (1987), they adopted $\rm H\alpha/H\beta = 3.1$ for active galaxies and 2.85 for \hii\ region galaxies (e.g., Ferland \& Netzer 1983; Gaskell 1984; Gaskell \& Ferland 1984),  and the large sample statistics suggested that the intrinsic value of broad-line $\rm H\alpha/H\beta $ is 3.06 with a standard deviation of 0.03 dex (Dong et al. 2008).
The steep Balmer decrement of 3C 445 suggests that 3C 455 is reddened, i.e., $E(B-V) \sim 1.23$ mag if the SMC extinction law is employed, which was  confirmed by the large $\rm Pa\alpha/H\beta$ ratio (5.6; Rudy \& Tokunaga 1982).
Furthermore, the Balmer decrement without the blue system contribution is $\rm H\alpha/H\beta\sim 8.0$ (Crenshaw et al. 1988), which predicts a slightly small reddening with $E(B-V) \sim 1.05$ mag. Indeed, it does not really matter what the value of the extinction exactly is, and the fact itself that the nuclear region of 3C 445 is heavily obscured is unquestionable.
In such a situation, the Ly$\alpha$ emission from the nuclear region would be completely absorbed, and it is not plausible in view of the detection of the Ly$\alpha$ line by Crenshaw et al. (1988). However, we find that the Ly$\alpha$ line is narrow and the profile of which is similar to those of \oiii\ $\rm \lambda\lambda4959,5007$ (see Fig. 1 in Crenshaw et al. 1988). The origin of the Ly$\alpha$ line is probable the star-forming in the host galaxy.
In order to check this guess, the broadband spectral energy distribution (SED) of 3C 445 taken with the SDSS (York et al. 2000), 2MASS (Skrutskie et al. 2006),  and \emph{WISE} (Wright et al. 2010) surveys are presented in Figure \ref{fig:f3}. Following the method of Zhang et al. (2017b), the broadband SED from the optical out to the middle-infrared can be modeled with the combination of a scaled and reddened quasar composite with $E(B-V)\sim 1.1$ mag and a scaled `$cst\_6Gyr\_z008$' template ($t = 6$ Gyr,  $Z =0.008$, and undergoing continuous star formation; Bruzual 2009) in the BC03 SSP library. These all are consistent with the Balmer decrement measurements.

3C 445 was  identified as a FR II radio source (Kronberg et al. 1986) with a round elliptical galaxy (Madrid et al. 2006) and a very bright unresolved Seyfert 1.5 nucleus (V{\'e}ron-Cetty \& V{\'e}ron 2010). 3C 445 is clearly lobe-dominated with a steep radio spectrum between 2.7 and 4.8 GHz ($\alpha^{4.8}_{2.7} = 0.7$) and a core-to-lobe intensity ratio of  $R = 0.039$ (Morganti et al. 1993). From the projected sizes of the supergalactic-scale biconical lobes, Eracleous \& Halpern (1998) inferred an inclination of $i > 60^\circ$. Moreover, Sambruna et al. (2007) derived an upper limit of the inclination angle $i < 71^\circ$ using the ratio of the radio fluxes of the approaching and receding jets ($\sim 7.7$; Leahy et al. 1997). These radio results suggest that 3C 445 is viewed at a very large observation angle ($60^\circ < i < 71^\circ$), which is consistent with the suggestion from the reddening properties. 3C 445 appears to host an obscured AGN. Indeed, the extremely red spectral energy distribution (Elvis et al. 1984; Crenshaw et al. 1988; Kotilainen et al. 1992), the large offsets of the hydrogen Balmer and Paschen lines relative to theoretical values (Osterbrock et al. 1976; Crenshaw et al. 1988;  Rudy \& Tokunaga 1982), and the X-ray continuums from the XMM-Newton, ROSAT and ASCA observations (Sambruna et al. 1998, 2007), consistently imply the nuclear region of 3C 445 obscured by the circumnuclear dust. Moreover, the polarization of the continuum (Brindle et al. 1990) with a trend of decreasing polarization degree with increasing wavelength (Rudy et al. 1983) also provides solid evidence for the presence of the  dust in 3C 445. Furthermore,  only the obscuring assembly in the AGN unified model, i.e., the dusty torus, can  provide such a large amount of dust. The observer's line of sight to 3C 445 would pass through the dusty torus. Based on the above properties, it can be concluded that 3C 445 is seen almost edge-on.

We suspect that the observation angle of view may be the key to  solving the mystery of the unique Balmer-line profiles in 3C 445. It is viewed at a large inclination angle that is actually conducive to the production of the double-peaked disk-line from the accretion disk and incidentally successfully explains the extreme reddening of the disk-line component and the red system of the Balmer lines. Moreover, the large  inclination angle also provides the possibility of the interaction between the outflowing winds and the peripheral expanding matters, e.g., the dusty gases in the torus. We acknowledge that 3C 445 is not a unique instance; the emission lines of 3C 445 show almost the same emission-line profiles as the other three cases (i.e., SDSS J132052.19+574737.3, SDSS J150718.10+312942.5 and SDSS 153636.22+044127.0) which we have reported.
In Zhang et al. (2019), the blue system of the emission-line profiles is considered to be the result of the shock-heated outflowing gases. Since the outflow can be accelerated to more than several thousand kilometers per second, it is easy to produce a high-speed and high-temperature ($T \ge 10^7 ~\rm K$) shock in the collision between the massive outflow and the inner surface of the dusty torus. Ionizing photons can be produced in the postshock plasma and diffuse upstream to form a photoionization front with a velocity  that exceeds the shock. Thus, the photoionization front would  be driven into the preshock high-density ($n\rm_H \ge 10^{12} ~ cm^{-3}$) gases in the torus, expanding as a precursor \hii\ region, and the fluxes of the blue system originating from the precursor \hii\ region ahead of the shock, which is also consist with the blue system having a more extreme reddening.
Additionally, in many quasars, the outflow gases, as the source of the shock, continuously flow outward from the inner region under the stable supply of the disk winds, but it is possible that the outflow in 3C 445   is discontinuous. In this case, the appearance/disappearance of the blue system (the result of the outflow) in 3C 445 is irregular, and the spectral archives just happened to catch the profile variability.

\section*{Acknowledgements}

This work is supported by National Natural Science Foundation of China (NSFC-11573024 and 11473025). This research has made use of the NASA/IPAC Extragalactic Data Base (NED), which is operated by the Jet Propulsion Laboratory, California Institute of Technology, under contract with NASA. Some of the data presented herein were obtained with the 120 inch telescope at Lick Observatory, the 1.8 m Perkins reflector at Lowell Observatory,  the 2.1 m telescope at the Kitt Peak National Observatory, the 4.2 m William Herschel Telescope at the Observatorio del Roque de los Muchachos, the 3.58 m optical/infrared Telescopio Nazionale Galileo, and the SDSS, 2MASS, and \emph{WISE} surveys.





\begin{thebibliography}{99}
\bibitem[Antonucci(1993)]{1993ARA&A..31..473A} Antonucci, R.\ 1993, \araa, 31, 473
\bibitem[Bruzual(2009)]{2009arXiv0911.0791B} Bruzual, A.~G.\ 2009, arXiv e-prints, arXiv:0911.0791
\bibitem[Buttiglione et al.(2009)]{2009A&A...495.1033B} Buttiglione, S., Capetti, A., Celotti, A., et al.\ 2009, \aap, 495, 1033
\bibitem[Boroson \& Lauer(2009)]{2009Natur.458...53B} Boroson, T.~A., \& Lauer, T.~R.\ 2009, \nat, 458, 53
\bibitem[Brindle et al.(1990)]{1990MNRAS.244..577B} Brindle, C., Hough, J.~H., Bailey, J.~A., et al.\ 1990, \mnras, 244, 577
\bibitem[Brocklehurst(1971)]{1971MNRAS.153..471B} Brocklehurst, M.\ 1971, \mnras, 153, 471
\bibitem[Chen et al.(1989)]{1989ApJ...339..742C} Chen, K., Halpern, J.~P., \& Filippenko, A.~V.\ 1989, \apj, 339, 742
\bibitem[Cohen et al.(1999)]{1999AJ....118.1963C} Cohen, M.~H., Ogle, P.~M., Tran, H.~D., et al.\ 1999, \aj, 118, 1963.
\bibitem[Collin-Souffrin et al.(1980)]{1980A&A....83..190C} Collin-Souffrin, S., Dumont, S., Heidmann, N., \& Joly, M.\ 1980, \aap, 83, 190
\bibitem[Corbett et al.(1998)]{1998MNRAS.296..721C} Corbett, E.~A., Robinson, A., Axon, D.~J., Young, S., \& Hough, J.~H.\ 1998, \mnras, 296, 721
\bibitem[Crenshaw et al.(1988)]{1988AJ.....96.1208C} Crenshaw, D.~M., Peterson, B.~M., \& Wagner, R.~M.\ 1988, \aj, 96, 1208
\bibitem[Decarli et al.(2010)]{2010ApJ...720L..93D} Decarli, R., Dotti, M., Montuori, C., Liimets, T., \& Ederoclite, A.\ 2010, \apjl, 720, L93
\bibitem[Dong et al.(2008)]{2008MNRAS.383..581D} Dong, X., Wang, T., Wang, J., et al.\ 2008, \mnras, 383, 581
\bibitem[Elvis et al.(1984)]{1984ApJ...280..574E} Elvis, M., Willner, S.~P., Fabbiano, G., et al.\ 1984, \apj, 280, 574
\bibitem[Eracleous \& Halpern(1993)]{1993ApJ...409..584E} Eracleous, M., \& Halpern, J.~P.\ 1993, \apj, 409, 584
\bibitem[Eracleous, \& Halpern(1994)]{1994ApJS...90....1E} Eracleous, M., \& Halpern, J.~P.\ 1994, \apjs, 90, 1
\bibitem[Eracleous \& Halpern(1998)]{1998ApJ...505..577E} Eracleous, M., \& Halpern, J.~P.\ 1998, \apj, 505, 577
\bibitem[Eracleous \& Halpern(2003)]{2003ApJ...599..886E} Eracleous, M., \& Halpern, J.~P.\ 2003, \apj, 599, 886
\bibitem[Ferland, \& Netzer(1983)]{1983ApJ...264..105F} Ferland, G.~J., \& Netzer, H.\ 1983, \apj, 264, 105
\bibitem[Gaskell(1984)]{1984ApL....24...43G} Gaskell, C.~M.\ 1984, Astrophysical Letters, 24, 43
\bibitem[Gaskell, \& Ferland(1984)]{1984PASP...96..393G} Gaskell, C.~M., \& Ferland, G.~J.\ 1984, \pasp, 96, 393
\bibitem[Halpern(1990)]{1990ApJ...365L..51H} Halpern, J.~P.\ 1990, \apjl, 365, L51
\bibitem[Jones et al.(2009)]{2009MNRAS.399..683J} Jones, D.~H., Read, M.~A., Saunders, W., et al.\ 2009, \mnras, 399, 683
\bibitem[Keel et al.(2005)]{2005ApJS..158..139K} Keel, W.~C., Irby, B.~K., May, A., et al.\ 2005, \apjs, 158, 139
\bibitem[Kotilainen et al.(1992)]{1992MNRAS.256..125K} Kotilainen, J.~K., Ward, M.~J., Boisson, C., et al.\ 1992, \mnras, 256, 125
\bibitem[Kronberg et al.(1986)]{1986A&A...169...63K} Kronberg, P.~P., Wielebinski, R., \& Graham, D.~A.\ 1986, \aap, 169, 63
\bibitem[Leahy et al.(1997)]{1997MNRAS.291...20L} Leahy, J.~P., Black, A.~R.~S., Dennett-Thorpe, J., et al.\ 1997, \mnras, 291, 20
\bibitem[Leighly(2004)]{2004ApJ...611..125L} Leighly, K.~M.\ 2004, \apj, 611, 125
\bibitem[Lewis et al.(2010)]{2010ApJS..187..416L} Lewis, K.~T., Eracleous, M., \& Storchi-Bergmann, T.\ 2010, \apjs, 187, 416
\bibitem[Li et al.(2015)]{2015ApJ...812...99L} Li, Z., Zhou, H., Hao, L., et al.\ 2015, \apj, 812, 99
\bibitem[Li et al.(2016)]{2016RAA....16..146L} Li, Z.-Z., Zhou, H.-Y., Hao, L., et al.\ 2016, Research in Astronomy and Astrophysics, 16, 146
\bibitem[Liu et al.(2016)]{2016ApJ...822...64L} Liu, W.-J., Zhou, H.-Y., Jiang, N., et al.\ 2016, \apj, 822, 64
\bibitem[Madrid et al.(2006)]{2006ApJS..164..307M} Madrid, J.~P., Chiaberge, M., Floyd, D., et al.\ 2006, \apjs, 164, 307
\bibitem[Marziani et al.(2013)]{2013ApJ...764..150M} Marziani, P., Sulentic, J.~W., Plauchu-Frayn, I., \& del Olmo, A.\ 2013, \apj, 764, 150
\bibitem[Morganti et al.(1993)]{1993MNRAS.263.1023M} Morganti, R., Killeen, N.~E.~B., \& Tadhunter, C.~N.\ 1993, \mnras, 263, 1023
\bibitem[Oke(1987)]{1987slrs.work..267O} Oke, J.~B.\ 1987, Superluminal Radio Sources, 267
\bibitem[Osterbrock et al.(1976)]{1976ApJ...206..898O} Osterbrock, D.~E., Koski, A.~T., \& Phillips, M.~M.\ 1976, \apj, 206, 898
\bibitem[Rudy et al.(1983)]{1983ApJ...271...59R} Rudy, R.~J., Schmidt, G.~D., Stockman, H.~S., \& Moore, R.~L.\ 1983, \apj, 271, 59
\bibitem[Rudy \& Tokunaga(1982)]{1982ApJ...256L...1R} Rudy, R.~J., \& Tokunaga, A.~T.\ 1982, \apjl, 256, L1
\bibitem[Sambruna et al.(2007)]{2007ApJ...665.1030S} Sambruna, R.~M., Reeves, J.~N., \& Braito, V.\ 2007, \apj, 665, 1030
\bibitem[Shapovalova et al.(2013)]{2013A&A...559A..10S} Shapovalova, A.~I., Popovi{\'c}, L.~{\v C}., Burenkov, A.~N., et al.\ 2013, \aap, 559, A10
\bibitem[Skrutskie et al.(2006)]{2006AJ....131.1163S} Skrutskie, M.~F., Cutri, R.~M., Stiening, R., et al.\ 2006, \aj, 131, 1163
\bibitem[Strateva et al.(2003)]{2003AJ....126.1720S} Strateva, I.~V., Strauss, M.~A., Hao, L., et al.\ 2003, \aj, 126, 1720
\bibitem[Strateva et al.(2004)]{2004ASPC..311..189S} Strateva, I., Strauss, M., Hao, L., et al.\ 2004, AGN Physics with the Sloan Digital Sky Survey, 311, 189
\bibitem[Tang \& Grindlay(2009)]{2009ApJ...704.1189T} Tang, S., \& Grindlay, J.\ 2009, \apj, 704, 1189
\bibitem[Veilleux, \& Osterbrock(1987)]{1987ApJS...63..295V} Veilleux, S., \& Osterbrock, D.~E.\ 1987, \apjs, 63, 295
\bibitem[V{\'e}ron-Cetty \& V{\'e}ron(2010)]{2010A&A...518A..10V} V{\'e}ron-Cetty, M.-P., \& V{\'e}ron, P.\ 2010, \aap, 518, A10
\bibitem[Wang et al.(2011)]{2011ApJ...738...85W} Wang, H., Wang, T., Zhou, H., et al.\ 2011, \apj, 738, 85
\bibitem[Wright et al.(2010)]{2010AJ....140.1868W} Wright, E.~L., Eisenhardt, P.~R.~M., Mainzer, A.~K., et al.\ 2010, \aj, 140, 1868
\bibitem[York et al.(2000)]{2000AJ....120.1579Y} York, D.~G., Adelman, J., Anderson, J.~E., Jr., et al.\ 2000, \aj, 120, 1579
\bibitem[Zhang(2013)]{2013MNRAS.431L.112Z} Zhang, X.-G.\ 2013, \mnras, 431, L112
\bibitem[Zhang et al.(2017)]{2017ApJ...836...86Z} Zhang, S., Zhou, H., Shi, X., et al.\ 2017a, \apj, 836, 86
\bibitem[Zhang et al.(2017)]{2017ApJ...845..126Z} Zhang, S., Zhou, H., Shi, X., et al.\ 2017b, \apj, 845, 126
\bibitem[Zhang et al.(2019)]{2019ApJ...877...33Z} Zhang, S., Zhou, H., Shi, X., et al.\ 2019, \apj, 877, 33
\end{thebibliography}





\bsp	
\label{lastpage}
\end{document}